\documentclass[fleqn,usenatbib]
{mnras}

\usepackage{newtxtext,newtxmath}

\usepackage[T1]{fontenc}
\usepackage{ae,aecompl}

\usepackage{graphicx}	
\usepackage{amsmath}	
\usepackage{amssymb}	

\title[Near-Infrared DIBs towards Her 36]{Near-Infrared Diffuse Interstellar Bands towards Her 36}

\author[M. G. Rawlings et al.]{
M. G. Rawlings,$^{1}$\thanks{E-mail: m.rawlings@eaobservatory.org (MGR)}
A. J. Adamson,$^{2}$
C. C. M. Marshall,$^{3}$
and P. J. Sarre$^{3}$
\\
$^{1}$East Asian Observatory / James Clerk Maxwell Telescope, 660 N. A'ohoku Place, Hilo, HI 96720, USA\\
$^{2}$Gemini Observatory, 670 N. A'ohoku Place, Hilo, Hawaii, 96720, USA\\
$^{3}$School of Chemistry, The University of Nottingham, University Park, Nottingham NG7 2RD, UK
}

\date{Accepted 2019 Mar 5. Received 2019 Fe 26; in original form 2018 May 29}

\pubyear{2019}

\begin{document}
\label{firstpage}
\pagerange{\pageref{firstpage}--\pageref{lastpage}}
\maketitle

\begin{abstract}
Discovered almost a century ago, the Diffuse Interstellar Bands (DIBs) still lack convincing and comprehensive identification. Hundreds of DIBs have now been observed in the near-ultraviolet (NUV), visible and near-infrared (NIR). They are widely held to be molecular in origin, and modelling of their band profiles offers powerful constraints on molecular constants. Herschel 36, the illuminating star of the Lagoon Nebula, has been shown to possess unusually broad and asymmetric DIB profiles in the visible, and is also bright enough for NIR observation. We present here high-resolution spectroscopic observations targeting the two best-known NIR DIBs at 11797.5 and 13175 \AA\ toward this object and a nearby comparison O-star, 9 Sgr, using the GNIRS instrument on Gemini North. We show a clear detection of the 13175 \AA\ DIB in both stars, and find (i) that it does not exhibit the unusual wing structure of some of the visual DIBs in Her 36 and (ii) that the depth of the band in the two objects is very similar, also contrary to the behaviour of the visual DIBs. We discuss the implications of these results for multiple DIB carrier candidates, and the location of their carriers along the observed lines of sight.
\end{abstract}




\section{Introduction}
\label{sec:Introduction}

\subsection{Background}
\label{subsec:Background}
The Diffuse Interstellar Bands (DIBs) are a major and long-standing challenge to modern day astronomical spectroscopy. They comprise approximately 500 interstellar absorption features predominantly in the visible (\citealt{2009ApJ...705...32H}; \citealt{2014IAUS..297...89Y};
\citealt{2016JPhCS.728f2005G}; \citealt{2018ApJS..237...40S}), and are observed towards stars suffering significant extinction from the diffuse interstellar medium. The DIB carriers are generally thought to be large, carbon-based, gas-phase molecules (see, e.g., \citealt{2006JMoSp.238....1S}; \citealt{2014IAUS..297.....C} and references therein for reviews), and with the exception of the very probable assignment of a few DIBs in the 9300 -- 9700 \AA\ region to the C$_{60}^{+}$ species based on helium-tagging experiments (\citealt{2015Natur.523..322C};
\citealt{2017ApJ...843L...2C}; \citealt{2017ApJ...846..168S}; \citealt{2017MNRAS.465.3956G}; \citealt{2017ApJ...843...56W}; \citealt{2018ApJ...858...36C}; 
\citealt{2018A&A...614A..28L}), the carriers of the DIBs are unknown.

Observations of DIB profiles and substructure offer a possible way to narrow the selection of possible carriers. Substructure has been known to exist in diffuse bands for a number of decades; since the first signs were seen by \citet{1982ApJ...252..610H}, many high- and ultra-high-resolution studies have been conducted on DIBs which display complex substructure (e.g. \citealt{1995MNRAS.277L..41S}; \citealt{1996MNRAS.283L.105K}; \citealt{1996A&A...307L..25E};\citealt{1998ApJ...495..941K}; \citealt{2002A&A...384..215G}; \citealt{2002A&A...396..987G}; \citealt{2008MNRAS.386.2003G}; \citealt{2008ApJ...682.1076G}). Modelling of marked asymmetries seen in some diffuse bands led \citet{2013ApJ...773...42O} to propose small, asymmetric, polar molecules as the carriers of some DIBs (see Section \ref{sec:Discussion}).

\subsection{DIBs in the Infrared}
\label{subsec:DIBsInTheInfrared}
Nearly 30 years ago, DIBs were first detected in the near-infrared (NIR). The first detections of bands in the \textit{J} window (1.2 $\mu$m) were by \citet{1990Natur.346..729J}, who detected two bands near 11797.5 and 13175 \AA. \citet{2014ApJ...796...58R} studied these two DIBs in many sightlines, and saw numerous cases of asymmetric band profiles; a common feature of strong DIBs in the visible. Band profile modelling showed that they could be fitted with two components of varying strength between sightlines. Other studies have now expanded the number of NIR DIBs, and a total of 15 features have now been detected (13 of which were identified by \citealt{2011Natur.479..200G}). These discoveries led to renewed interest in IR DIBs. \cite{2014A&A...569A.117C} obtained medium-resolution spectra of eight established DIBs targets spanning 3000 to 25000~\AA, confirming the presence of nine of the original thirteen NIR DIBs candidates of \cite{2011Natur.479..200G}, as well as identifying an additional eleven DIB candidates. In recent years, very large spectroscopic datasets have also allowed statistical evaluations to be made regarding DIB properties for the first time. \cite{2013ApJ...778...86K} carefully combined spectra from dozens of radial velocity survey sources to perform a detailed analysis of the DIB at 8620~\AA. The careful extraction of DIB spectra by \cite{2015ApJ...798...35Z} from a very large number of low-extinction APOGEE \citep{2013AJ....146...81Z} survey sources demonstrated the possibility of DIB mapping on large scales and analyses involving large statistical samples. This gave rise to a number of follow-up studies by (e.g.) \citet{2017MNRAS.467.3099G}, \citet{2017A&A...600A.129E} and \cite{2016ApJ...821...42H} and references therein.

The interest in infrared DIBs arises because of their potential for identification with some specific carriers with transitions in the NIR. The initial paper by \citet{1990Natur.346..729J} noted that ionized C$_{60}$ had transitions at approximately the wavelength of the observed bands; subsequently other possibilities have been raised, including some of the larger Polycyclic Aromatic Hydrocarbon (PAH) ions \citep{2005ApJ...629.1188M}. 

Further study of these bands could therefore provide additional clues to the nature of the DIB carriers, if those carriers could also be demonstrated to give rise to other bands.

\subsection{Herschel 36}
\label{subsec:Herschel36}
Herschel 36 (hereafter Her 36) is the ionisation source of the bright part of the Lagoon nebula (M8).  A multiband near-IR image of Her 36 and the surrounding nebula can be seen in Fig. 1 of \citet{2006ApJ...649..299G}, which reveals emission at 2.17 and 3.8 $\mu$m. The star is also bright in the visible; more recent reports show that this sightline exhibits strong diffuse bands at visible wavelengths \citep{2013ApJ...773...41D}. These bands are uniquely broader and more asymmetric than those seen anywhere else (see e.g. fig. 6 of \citealt{2013ApJ...773...41D}). Although some interstellar atomic lines in this sightline appear to show only a single Doppler component (e.g. \citealt{2018A&A...615A.158T}), \citet{2017A&A...604A.135D} found that sodium D lines exhibited multiple velocity components in the region. Nevertheless,  it appears that the extreme DIB asymmetry seen in Her 36 is very localized, and hence probably intrinsic to the bands arising in its immediate environment. This sightline therefore offers the opportunity to study both DIB asymmetry and the long-wavelength extreme of the known DIB spectrum at the same time.

\citet{2013ApJ...773...41D} noted that approximately 1/3 of the interstellar colour excess seen toward Her 36 may arise in a known foreground HI self absorption (HISA) cloud \citep{1972A&A....18...55R, 1984NASCP2345..117C}. This cloud, known as the Riegel-Crutcher (R-C) cloud, is at a distance of approximately 120 -- 150 pc. Indeed, \citet{2013ApJ...773...41D} speculated that some of the weaker DIBs they detected toward Her 36 might be entirely attributed to this foreground cloud, rather than interstellar material local to Her 36.

The production of anomalous visual-wavelength DIB profiles in the particular radiation environment of Her 36 combined with the comparatively small number of proposed carrier types of the NIR DIBs led us to attempt to determine if the NIR DIBs (specifically the two observed by \citealt{1990Natur.346..729J}) also show the anomalous red wings seen in the visual (or, indeed, if the carriers are present at all). We present here high-resolution spectroscopic observations of Her 36 centred on the wavelengths of the two best-known NIR diffuse bands, at 11797.5 and 13175 \AA, with sufficiently high s/n ratio to detect the NIR DIBs and clearly detect any asymmetries comparable to those seen in the visual band DIBs.

\section{Observations and Data Reduction}
\label{sec:ObservationsAndDataReduction}
$J$-band spectroscopy was conducted using the Gemini North telescope and the GNIRS instrument in high-resolution mode (110.5 l/mm grating and long camera). The observations were taken on the night of 2015 June 02 (UTC) under the project code GN-2015A-FT-22. Observations were made of both the science field, Her 36 (O7.5 V; $J = 7.938$, $H = 7.450$, $K = 6.911$; $E_{B-V} = 0.87$) and a bright calibration standard, 9 Sgr (O4 V; $J = 5.80$, $H = 5.78$, $K = 5.84$; $E_{B-V} = 0.33$). The science target position observed was at 18:03:40.330 -24:22:42.708 (J2000) with a slit position angle of $36^{\circ}$ East of North, chosen to avoid nearby faint field stars on the slit by reference to the VLT image by \citet{2006ApJ...649..299G}. All science observations made use of a 3-arcsec nod, incorporating frames of a suitable off-source reference position outside any extended Her 36 emission.  Flat field frames were taken to enable the usual corrections to be applied. XeAr arc lamp observations suitable for use at the target wavelengths (11797.5 and 13175 \AA) were also taken, enabling wavelength calibration of both the science and calibration targets.

The data were reduced using standard techniques via a combination of the GNIRS/IRAF \citep{2005ASPC..347..514C} and Starlink \citep{2014ASPC..485..391C} data reduction software suites. For both targets, the groups of observing frames were co-added, cosmic ray spikes were removed, and the nodded pairs of spectra extracted and then co-added to generate one-dimensional spectra. These spectra were wavelength-calibrated against arc line spectra. 

The telluric standard, a binary O star within a few arcminutes of Her 36, has a not-insignificant degree of reddening (E(B-V)=0.33). Initially, it appeared that the  standard star spectra exhibited no evidence of DIBs at the target wavelengths. However, the presence of strong telluric absorptions combined with the relatively weak predicted DIB strengths led the referee to suggest that we employ an atmospheric modelling approach to check the telluric standard for DIB features. We therefore downloaded simulated spectra from the TAPAS online telluric modelling facility \citep{2014A&A...564A..46B} and ratioed the telluric standard to these spectra. The results were clear: the chosen telluric standard standard has a DIB feature near 1.32 $\mu$m, and is therefore not a useful telluric ratioing star. However, the TAPAS spectra reproduce the atmosphere above Maunakea so well (for the night in question) that they can be used to remove telluric features (at least for the broader 13175~\AA\ band).

\subsection{Spectra}
\label{subsec:Spectra}
\begin{figure}
	{\includegraphics[width=\columnwidth, angle=270, scale=0.75]{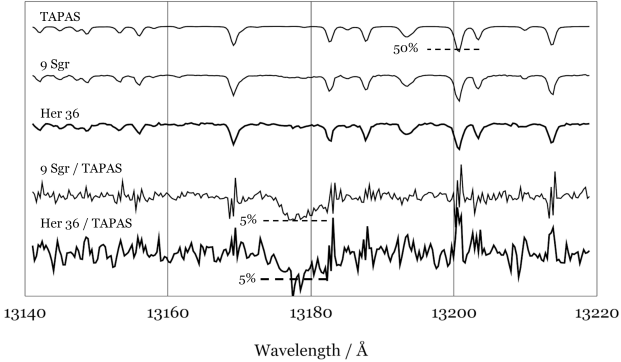}}
    \caption{The 13175~\AA\ DIB in Her 36 and 9 Sgr. Top to Bottom: TAPAS spectrum, 9 Sgr, Her 36, 9 Sgr/TAPAS, Her 36/TAPAS. The bottom two spectra are on  a  ten-times multiplied scale. Dashed lines indicate a 5 per cent feature depth. The top three spectra are plotted unscaled, and shifted for clarity. a dashed line below the TAPAS spectrum shows the 50 per cent depth of the deepest atmospheric feature.}
    \label{fig:Fig1}
\end{figure}

Fig. \ref{fig:Fig1} shows the 1.32 $\mu$m DIB region. The ratios in the lower part of the plot are expanded by a factor of 10 vertically to better display the diffuse band. The band is clearly present in both Her 36 and 9 Sgr. The unratioed spectra of both objects and the TAPAS spectrum used to remove telluric features are plotted at the top. Given the noise level in the Her 36 ratioed spectrum, we do not attempt detailed profile fitting. Instead we fit a single Gaussian with FWHM 5 \AA\ and one free parameter (the central depth). This yields a central depth of $5.5 \pm 0.7$ per cent and for a band with the profile seen in the diffuse ISM \citep{2014ApJ...796...58R} an equivalent width of $(2.96 \pm 0.4)\times 10^{-5}\ \mu$m. The central wavelength of the feature is approximately 3 \AA\ longward of the canonical 13175 \AA, in both Her 36 and 9 Sgr. 

\begin{figure}
	{\includegraphics[width=\columnwidth, angle=270, scale=0.75]{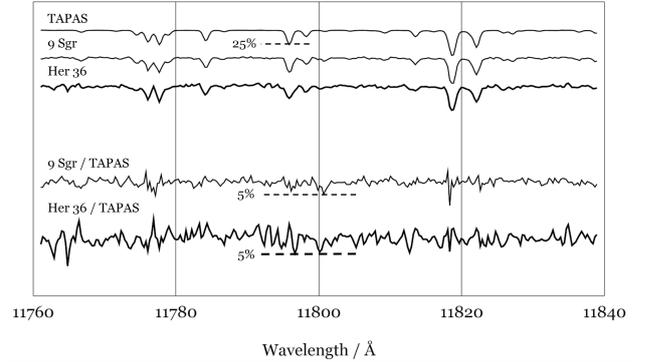}}
    \caption{The 11797.5~\AA\  DIB. Spectra are plotted in the same order as in Fig. \ref{fig:Fig1}, but with  magnification (five times) in the y axis for the lower two traces. Dashed lines again show representative depths below the continuum.}
    \label{fig:Fig2}
\end{figure}

Fig. \ref{fig:Fig2} shows the 1.17 $\mu$m region. There is a hint of a feature at approximately 1.18 $\mu$m, stronger in 9 Sgr than in Her 36. If this is the 11797.5 \AA\ DIB, then (i) its wavelength is again slightly longer than canonical and (ii) its depth is about the same as that of the band in the 1.32 $\mu$m region. The formal best fit to this redward feature, using a single Gaussian of FWHM 2 \AA, has a central depth of $3 \pm 0.5$ per cent, and for a feature of the same profile as in the diffuse ISM an equivalent width of $(0.63 \pm 0.1) \times 10^{-5}\ \mu$m. If we assume the feature is instead at the nominal wavelength of 11797.5 \AA\, then it is undetected ($1.1 \pm 1$ per cent in depth). The noise level is high enough, particularly in Her 36, that we do not discuss this spectral region further here.

\section{Discussion}
\label{sec:Discussion}

The near-IR DIB close to 1.32 $\mu$m is detected in Her 36 and the binary O star 9 Sgr located approximately two arcminutes to its northeast. The band is well resolved in our data and has the same central depth (approximately 5 per cent) in both lines of sight, although the selective extinction to Her 36 is greater by a factor of three. Both lines of sight show very similar band profiles - the DIB is neither significantly broadened nor more asymmetric than it is in the diffuse ISM (e.g. \citealt{2014ApJ...796...58R};  \citealt{2011Natur.479..200G} and references therein). This is in striking contrast to the behaviour of the visual DIBs (5780, 5797, 6196, 6613) \citep{2013ApJ...773...41D}. The band in Her 36 is close to the depth seen at similar extinction in the diffuse ISM \citep{2014ApJ...796...58R}. In 9 Sgr, it is a factor of at least two stronger than the diffuse ISM value. This is well within the known variability between lines of sight for visual DIBs, which can approach a factor of 10 \citep{2017ApJ...850..194F}, but very different from the behaviour of the visual DIBs referred to above, all of which are weaker in 9 Sgr than in Her 36 \citep{2013ApJ...773...41D}. 

Both stars lie behind a large, cold foreground cloud (the R-C cloud), which may provide up to 1/3 of the reddening to Her 36 \citep{2013ApJ...773...41D}. Radio measurements show a filamentary structure (e.g. \citealt{2018MNRAS.479.1465D}), but lack the spatial resolution to determine whether the foreground component of extinction could differ significantly for the two stars. The remainder of the extinction to Her 36 appears to be produced in material which is local to it.

The similarity in band profile and difference in depth per unit total extinction may arise in one of three ways. Firstly (and most simply), the band may arise entirely in the foreground cloud. If so, the material in the immediate  environs of Her 36 itself is highly inefficient at producing the 13175~\AA\ DIB. Secondly, 9 Sgr could be seen through a filament in the foreground cloud while Her 36 is seen through a transparent region. In this scenario, all the extinction seen toward Her 36 would be local to it. The local extinction would need to weaken the band strength while not altering the profile. Thirdly, as described by \citet{2017A&A...604A.135D} and \citet{2018A&A...615A.158T}, Her 36 is seen through both its local HII region and the surrounding PDR, while 9 Sgr is seen only through the PDR. Ignoring the foreground contribution in this case, the differences between the material in the PDR and that in the HII region would then need to drive the difference in the band strength. Although some previous observations of the visual DIBs toward 9 Sgr (e.g. \citealt{2017ApJ...835..107X} and references therein) have been conducted, further direct comparative studies of the visual (and NIR) DIBs in 9 Sgr would be helpful in resolving these differences.

The infrared DIB near 1.32$~\mu$m in Her 36 and 9 Sgr behaves differently in both strength and band profile from the four strong DIBs (at 5780, 5797, 6196 and 6613 \AA) in the visual. In all the scenarios described above, these differences must be driven in some way by the properties of the material immediately adjacent to Her 36, which is presumably modified by the local radiation environment and whose physical state apparently produces very extended red wings on some of the visual DIBs.

The absence or extreme weakness of the 11797.5~\AA\ feature compared with 13175~\AA\ towards Her 36 is not unexpected, as the equivalent width ratio between these bands ranges from 3 to 5.5 over a selection of targets \citep{2014ApJ...796...58R}, making detectability of the weaker band challenging with the current signal-to-noise ratio. The lack of a significant extended wing in the 13175-\AA\ DIB towards Her 36 may be due to the nature of the absorber; not all diffuse bands in the visible show such extended red wings.  This may be due to the carrier being non-polar \citep{2013ApJ...773...41D},  or lacking low-lying vibronic states \citep{2015MNRAS.453.3912M}.
Disentanglement of the relative contributions of foreground and local material to the observed DIB profiles towards Her 36 and 9 Sgr is a major challenge. This was discussed by \citet{2013ApJ...773...42O} in their figure 8, in which two possible scenarios for four of the visible DIBs are shown. It would be of interest to obtain near-IR spectra of higher signal-to-noise and at higher spectroscopic resolution to assist in separation of the relative contributions of the foreground cloud and the region local to Her 36.

If the carrier of the infrared DIB near 1.32$~\mu$m is indeed C$_{60}^{+}$ (or a similar species), and if the unusual visual bands arise due to the presence of small organic molecules (such as those discussed by \citealt{2013ApJ...773...42O}), then the immediate environment of Her 36 must somehow affect the small molecules in such a way as to produce the extended visual DIB red wings, while simultaneously affecting the (abundance of?) the C$_{60}^{+}$ species in such a way as to weaken the NIR DIB.

\section*{Acknowledgements}
Based on observations obtained at the Gemini Observatory, which is operated by the Association of Universities for Research in Astronomy, Inc., under a cooperative agreement with the NSF on behalf of the Gemini partnership: the National Science Foundation (United States), the National Research Council (Canada), CONICYT (Chile), Ministerio de Ciencia, Tecnolog\'{i}a e Innovaci\'{o}n Productiva (Argentina), and Minist\'{e}rio da Ci\^{e}ncia, Tecnologia e Inova\c{c}\~{a}o (Brazil). The presented data were partially processed using the Gemini IRAF package. The Starlink software \citep{2014ASPC..485..391C} is currently supported by the East Asian Observatory. MGR wishes to thank the East Asian Observatory for supporting this research. PJS thanks the Leverhulme Trust for award of a Leverhulme Emeritus Fellowship. C. C. M. M. thanks EPSRC for financial support and the University of Nottingham for a BESTS prize award.  The authors wish to thank Jan Cami for his comments and advice regarding DIB observations towards Her 36. The authors would also like to particularly thank the Referee, Giacomo Mulas, for his constructive feedback and (in particular) for his very helpful suggestion on telluric corrections: these resulted in changes that significantly improved this paper. Finally, this research made use of the TAPAS online atmospheric transmission modelling service, operated at the Institut Pierre Simon Laplace, France. 





\bibliographystyle{mnras}
\bibliography{DIBs_MGR_bibtex}
%

\bsp	
\label{lastpage}
\end{document}